\documentclass[A4paper]{jpconf}
\usepackage[dvips]{color}
\usepackage{graphicx}
\usepackage{bm}

\renewcommand{\d}{\mathrm{d}}

\newcommand{\be}{\begin{equation}}
\newcommand{\ee}{\end{equation}}
\newcommand{\bea}{\begin{eqnarray}}
\newcommand{\eea}{\end{eqnarray}}
\newcommand{\bse}{\begin{subequations}}
\newcommand{\ese}{\end{subequations}}

\newcommand{\pf}{k_{\mathrm F}}
\newcommand{\kf}{k_{\mathrm F}}
\newcommand{\ef}{\varepsilon_{\mathrm F}}

\newcommand{\q}{q}

\begin{document}
\nocite{*}

\title{Approximate expression for the dynamic structure factor in the Lieb-Liniger model}


\author{Alexander~Yu~Cherny$^\dag$ and Joachim~Brand$^\ddag$}
\address{$^\dag$Bogoliubov Laboratory of Theoretical Physics, Joint Institute for Nuclear
Research, 141980, Dubna, Moscow region, Russia}
\address{$^\ddag$Centre for Theoretical Chemistry and Physics and Institute of Fundamental
Sciences, Massey University, Private Bag 102~904, NSMC, Auckland, New Zealand}
\eads{\mailto{cherny@theor.jinr.ru}}

\begin{abstract}
Recently, Imambekov and Glazman [{\it Phys.~Rev.~Lett.}~{\bf 100}, 206805 (2008)] showed 
that the dynamic structure factor (DSF) of the 1D Bose gas demonstrates power-law 
behaviour along the limiting dispersion curve of the collective modes and calculated the 
corresponding exponents exactly. Combining these recent results with a previously 
obtained strong-coupling expansion we present an interpolation formula for the DSF of 
the 1D Bose gas. The obtained expression is further consistent with exact low energy 
exponents from Luttinger liquid theory and shows nice agreement with recent numerical 
results.
\end{abstract}

Cigar-shaped traps with cold alkali atoms have recently been used to obtain a quasi-1D 
quantum degenerate Bose gas, where atomic motion in the transverse dimensions is 
confined to zero-point quantum oscillations, in weak and strong interaction 
regimes~\cite{goerlitz01,weiss04}. Theoretically, we may describe the system as a 
one-dimensional rarefied gas where interactions of bosonic atoms can be described well 
by effective $\delta$-function interactions \cite{olshanii98}. Thus the Lieb-Liniger 
model \cite{lieb63:1,lieb63:2} is applicable. Being exactly solvable in the uniform 
case, the model, however, does not admit complete analytic solutions for the correlation 
functions. Up to now, this has been an outstanding problem in 1D 
physics~\cite{korepin93:book,giamarchi04:book}. Here, we propose an approximate formula 
for the DSF of the Lieb-Liniger gas that is consistent with known results in accessible 
limits and power laws.

Dynamical density-density correlations, which can be measured by the two-photon Bragg 
scattering \cite{stenger99,ozeri04}, are described by the {\em dynamic structure factor} 
(DSF) \cite{pitaevskii03:book}
\begin{equation}
S(q,\omega)=L\int \frac{\d t\d x}{2\pi\hbar}\,e^{i(\omega t-q x)}
\langle0|\delta\hat{\rho}(x,t)\delta\hat{\rho}(0,0)|0\rangle. \label{eqn:dsfdef}
\end{equation}
Here, we introduce the density fluctuations 
$\delta\hat{\rho}(x,t)\equiv\hat{\rho}(x,t)-n$ and the equilibrium density of particles 
$n=N/L$. We consider the case of zero temperature, where  $\langle0|\ldots|0\rangle$  
means ground-state average. The DSF is proportional to the probability of exciting the 
collective mode from the ground state with momentum $\q$ and energy $\hbar\omega$ 
transfer, as one can see in the energy representation of Eq.~(\ref{eqn:dsfdef})
\begin{equation}
S(k,\omega)=\sum_n |\langle0|\delta\hat{\rho}_k|n\rangle|^2\delta(\hbar\omega-E_n+E_0),
\label{eqn:dsfenergy}
\end{equation}
where $\delta\hat{\rho}_k=\sum_{j}e^{-i k x_j}$ is the Fourier component of
$\delta\hat{\rho}(x)$.

The Lieb-Liniger model \cite{lieb63:1,lieb63:2} represents a uniform 1D system of 
spinless bosons of mass $m$, interacting with pairwise point interactions $V(x)= g_{\rm 
B}\delta(x)$; the interaction strength $g_{\rm B}$ is assumed to be positive. Periodic 
boundary conditions are imposed on the wave functions. The strength of interactions can 
be measured in terms of the dimensionless Lieb-Liniger parameter $\gamma\equiv
 m g_{\mathrm{B}}/(\hbar^2 n)$. Within the Lieb-Liniger model, the DSF has the following
well-established properties.

i) Luttinger liquid theory predicts a power-law behaviour of the DSF at low energies in 
the vicinity of the momenta $k=0,2\pi n, 4\pi n\ldots$ and yields model-independent 
values of the exponents \cite{haldane81,astrakharchik04}. In particular, one can show 
\cite{astrakharchik04,castro_neto94} that in the vicinity of ``umklapp" point ($k=2\pi 
n$, $\omega =0$)
\begin{equation}
\label{pitdsf}
S(k,\omega)\sim (\omega^{2}-\omega^{2}_{-})^{K-1},
\end{equation}
where $K\equiv \hbar\pi n/(m c)$ and $c$ is sound velocity. Furthermore, within the 
Luttinger-liquid theory, the dispersion is linear in vicinity of the umklapp point: 
$\omega_{-}(k)\simeq c|k-2 \pi n|$. Relation (\ref{pitdsf}) leads to different exponents 
precisely at the umklapp point and outside of it:
\begin{equation}
\label{pitdsfexp}
S(k,\omega)\sim \left\{\begin{array}{ll}
\omega^{2(K-1)},& k=2\pi n,\\
(\omega-\omega_{-})^{K-1},& k\not=2\pi n.
\end{array}\right.
\end{equation}

ii) By using in a non-trivial manner the Bose-Fermi mapping in 1D \cite{cheon99}, the 
authors developed the time-dependent Hartree-Fock scheme \cite{brand05,cherny06} in the 
strong-coupling regime with the small parameter $1/\gamma$. The scheme guarantees 
validity of the DSF expansion \cite{brand05,cherny06}
\begin{equation}
S(k,\omega)\frac{\ef}{N}= \frac{\kf}{4 k}\left(1+\frac{8}{\gamma}\right)
+\frac{1}{2\gamma}\ln \frac{\omega^{2}-\omega_{-}^{2}} {\omega_{+}^{2}-\omega^{2}}+
O\left(\frac{1}{\gamma^2}\right),
\label{DSFlinear}
\end{equation}
for $\omega_{-}\leq\omega\leq\omega_{+}$, and zero otherwise. Here $\omega_\pm(k)$ are 
the limiting dispersions that bounds quasiparticle-quasihole excitations 
\cite{lieb63:2}. In the strong-coupling regime they take the form $\omega_\pm(k)= {\hbar 
|2 \pf k \pm k^2|}(1-4/\gamma)/{(2 m)}+ O\left(\gamma^{-2}\right)$. By definition, 
$\kf\equiv\pi n$ and $\ef\equiv\hbar^{2}\kf^{2}/(2m)$ are the Fermi wave vector and 
energy of a non-interacting Fermi gas, respectively.

\begin{figure}[b]
\begin{minipage}{.45\textwidth}
\includegraphics[width=1.07\textwidth]{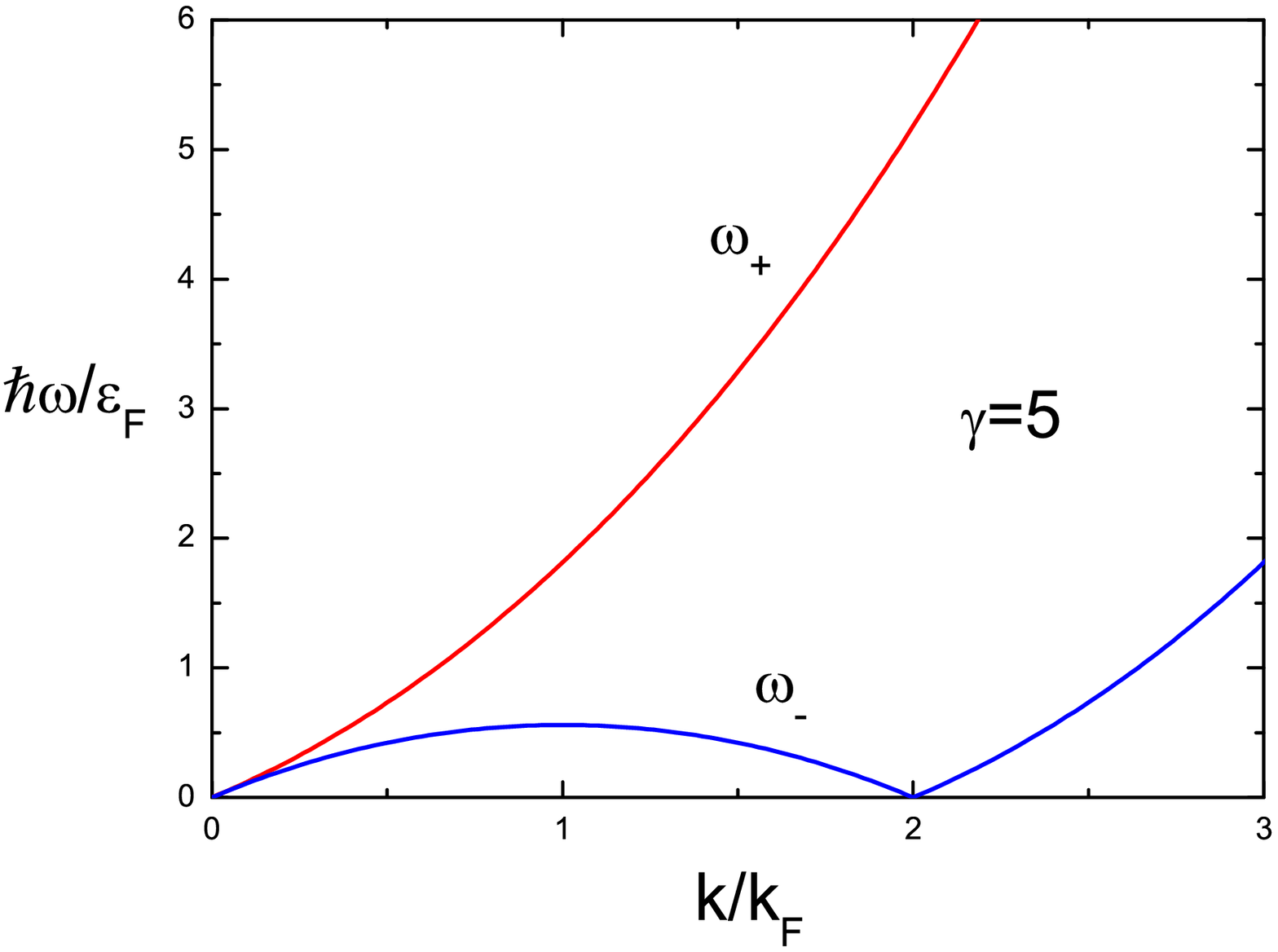}
\caption{\label{fig:omplmi} Limiting dispersions $\omega_{\pm}$ versus wave vector $k$ 
for the coupling parameter $\gamma=5$. The data are obtained numerically by solving 
Lieb-Liniger's system of integral equations \cite{lieb63:2,korepin93:book}. }
\end{minipage}\hspace{.1\textwidth}%
\begin{minipage}{.45\textwidth}
\includegraphics[width=\textwidth]{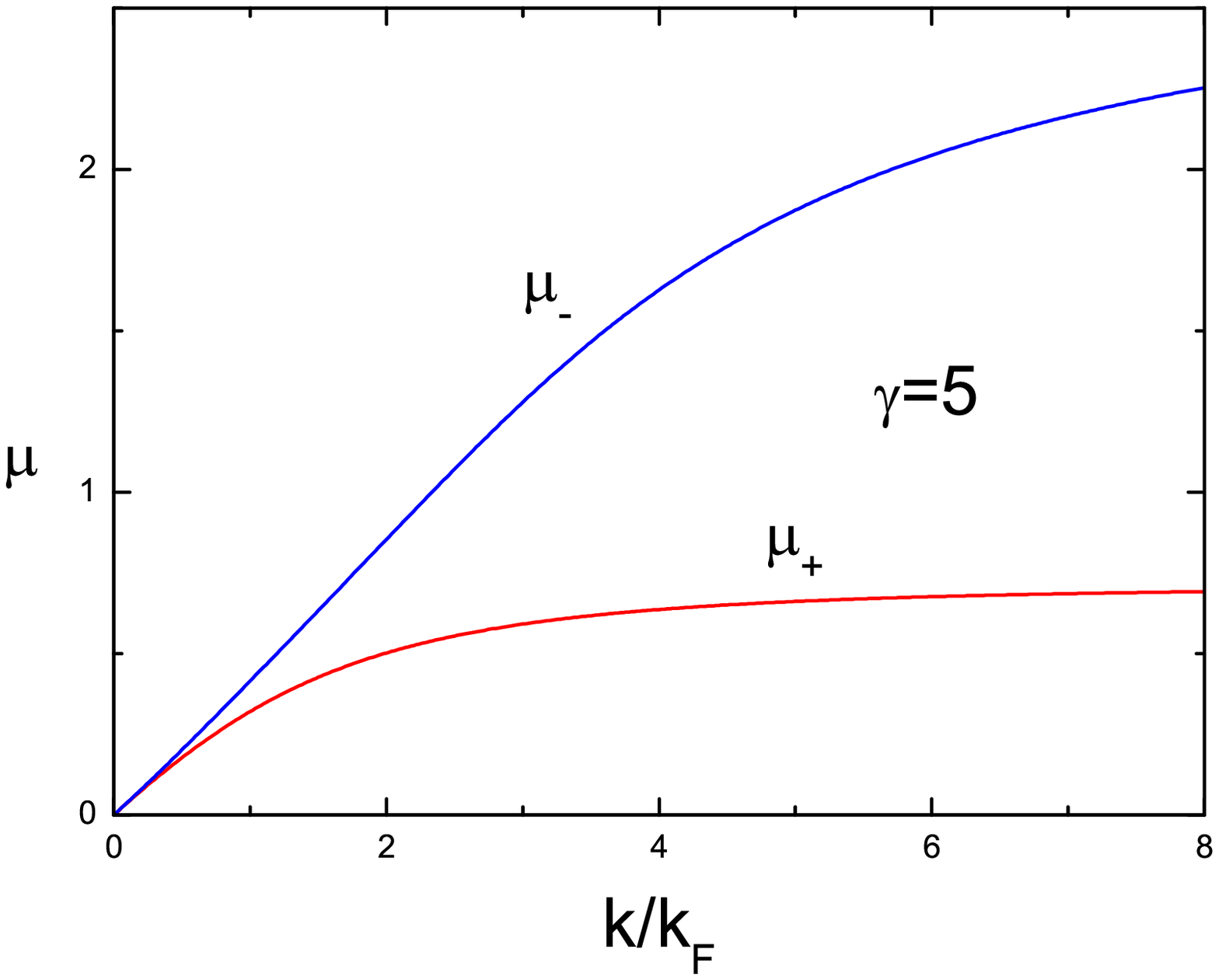}
\caption{\label{muplmi} Typical behaviour of the exact exponents in Eq.~(\ref{glazexp}). 
The plot shows $\mu_\pm$ for $\gamma=5$ obtained numerically using the method of 
Ref.~\cite{imambekov08}. }
\end{minipage}
\end{figure}

iii) As was shown by Imambekov and Glazman \cite{imambekov08}, in the Lieb-Liniger model
the DSF demonstrates power-law behaviour near the borders $\omega_\pm(k)$
\begin{equation}
S(k,\omega)\sim \big|\omega-\omega_{\pm}(k)\big|^{\mp \mu_{\pm}(k)}.
\label{glazexp}
\end{equation}
The positive exponents $\mu_\pm$ \cite{note} are related to the quasi-particle scattering 
phase and can be easily evaluated by solving a system of a few integral equations in 
thermodynamic limit \cite{imambekov08}. We obtain the exact relation
\begin{equation}
\label{mumiumklapp}
\mu_{-}(2\pi n-0)=2\sqrt{K}(\sqrt{K}-1),
\end{equation}
which obviously differs from the Luttinger liquid exponent (\ref{pitdsfexp}) for 
$k\not=2\pi n$. However, Imambekov's and Glazman's result (\ref{mumiumklapp}) is correct 
in the immediate vicinity of $\omega_\pm$ provided that the finite curvature of 
$\omega_{-}(k)$ is taken into consideration. Thus the difference in the exponents can be 
treated \cite{imambekov08} as an artifact of the linear spectrum approximation in the 
Luttinger liquid theory. Note, however, that the thin ``strip" in $\omega$-$k$ plane 
where the exponents are different vanishes in the point $k=2\pi n$; hence, the Luttinger 
exponent $2(K-1)$ should be exact here.

iv) The DSF can be calculated numerically by means of algebraic Bethe ansatz \cite{caux06}.

\begin{figure}[t]
\begin{center}
\includegraphics[width=.49\textwidth]{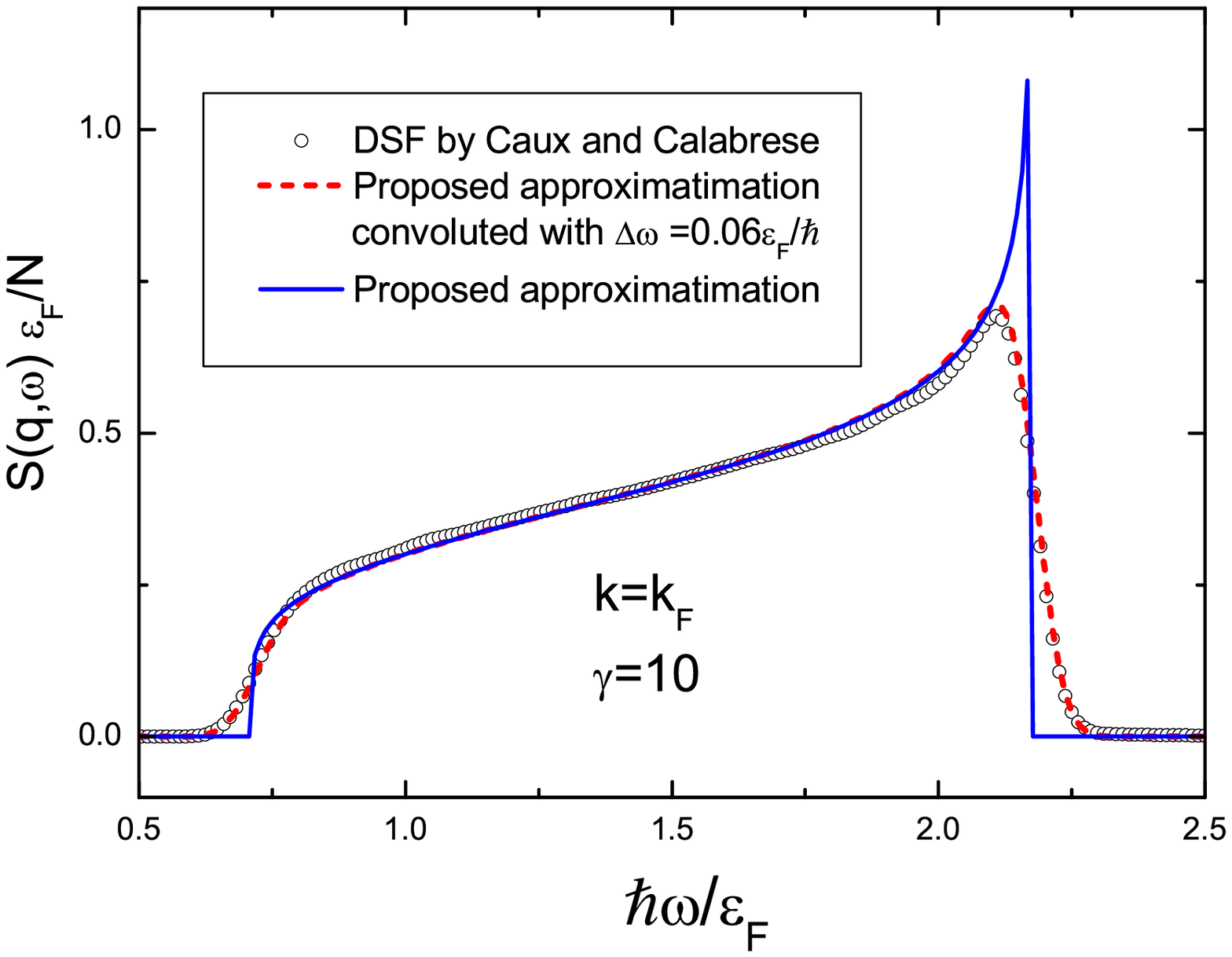}
\includegraphics[width=.49\textwidth]{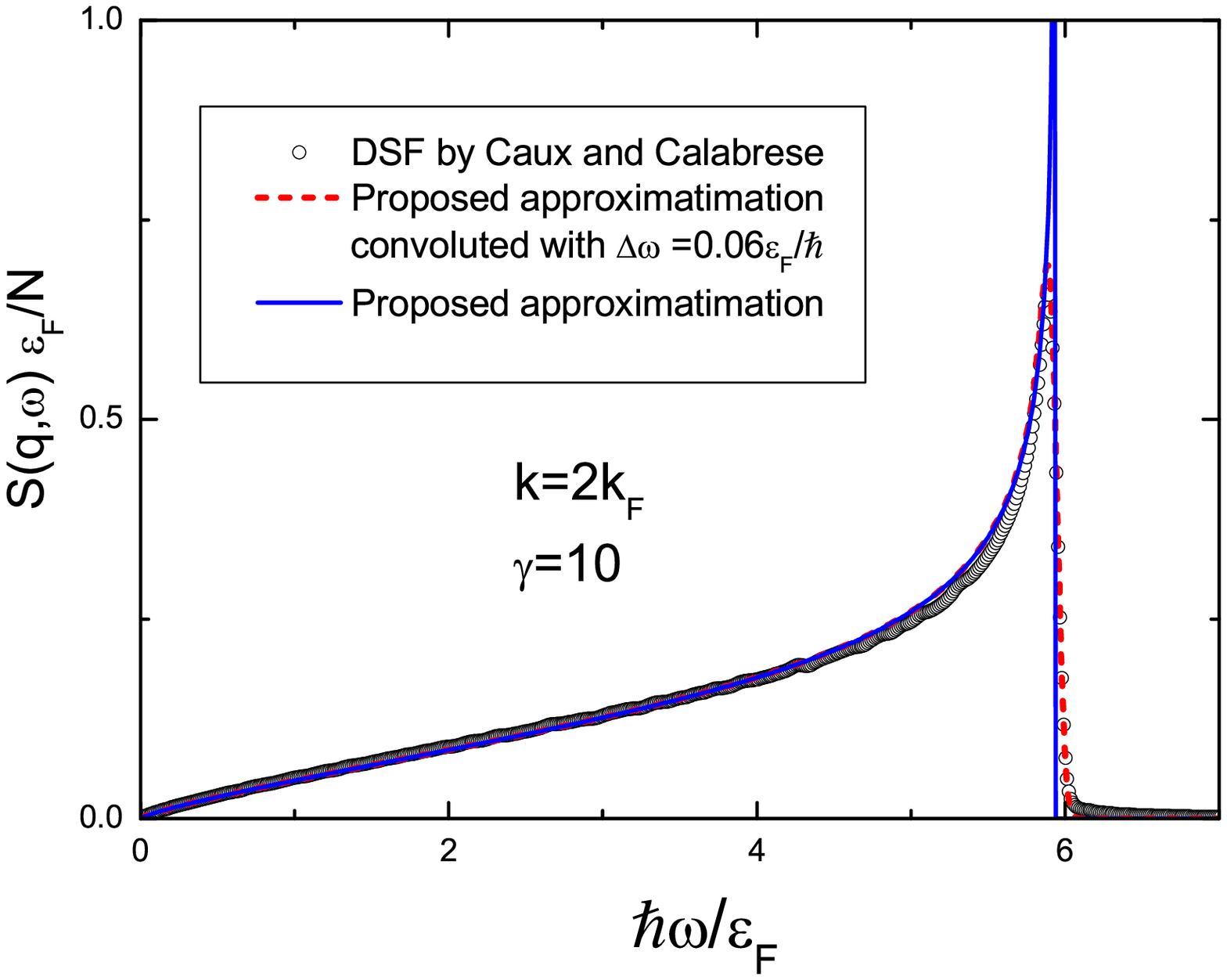}
\end{center}
\caption{\label{fig:dsf} The Dynamic Structure Factor (DSF) in the thermodynamic limit. 
The proposed approximation (\ref{dsfapp1}) (line) is compared to numerical data from Caux 
and Calabrese \cite{caux06} (open dots). The dashed (red) line shows the data of 
Eq.~(\ref{dsfapp1}) convoluted in frequency with a Gaussian of width 
$\Delta\omega\sqrt{2\ln2}=0.07\ef/\hbar$ in order to simulate smearing that has been 
used in generating the numerical results of Ref.~\cite{caux06}. The numerical data of 
Ref.~\cite{caux06} suggests that contributions from multi-particle excitations for 
$\omega>\omega_+$ (sharp line in parts a and b) are very small. Such contributions are 
not accounted for by the formula (\ref{dsfapp1}). }
\end{figure}

Here we suggest a phenomenological expression, which is consistent with all of the 
above-mentioned results. It reads
\begin{equation}
\label{dsfapp1}
S(k,\omega)=C \frac{(\omega^{\alpha}-\omega_{-}^{\alpha})^{\mu_{-}}}
{(\omega_{+}^{\alpha}-\omega^{\alpha})^{\mu_{+}}}
\end{equation}
for $\omega_{-}(k)\leq\omega\leq\omega_{+}(k)$, and zero otherwise. Here $C$ is a 
normalization constant, $\mu_{+}(k)$ and $\mu_{-}(k)$ are the exponents of 
Eq.~(\ref{glazexp}), and $\alpha\equiv 1+1/\sqrt{K}$. The normalization constant depends 
on momentum but not frequency and can be determined from the $f$-sum rule (see, e.g., 
Ref.~\cite{pitaevskii03:book})
\begin{equation}
\int_{-\infty}^{+\infty} \d\omega\, \omega S(q,\omega)= N\frac{q^{2}}{2m}. \label{fsum}
\end{equation}
We assume that in Eq.~(\ref{dsfapp1}) the value of the exponent $\mu_{-}(k=2\pi n)$ 
coincides with its limiting value (\ref{mumiumklapp}) in vicinity of the umklapp point.

Now it can be easily seen from (\ref{dsfapp1}) that
\begin{equation}
\label{glazdsfexp} S(k,\omega)\sim \left\{\begin{array}{ll}
\omega^{2(K-1)},& k=2\pi n,\\
(\omega-\omega_{-})^{\mu_{-}(k)},& k\not=2\pi n.
\end{array}\right.
\end{equation}
Thus, the suggested formula is consistent with the both the Luttinger liquid behaviour at 
the umklapp point and Imambekov's and Glazman's power-law behaviour in vicinity of it, as 
it should be.

In the strong-coupling regime, Eq.~(\ref{dsfapp1}) correctly yields the first order 
expansion (\ref{DSFlinear}). In order to prove this, it is sufficient to use the 
strong-coupling values of $K=1 +4/\gamma + O(1/\gamma^2)$, 
$\mu_{\pm}(k)=2\arctan[k/(n\gamma)]/\pi+ O(1/\gamma^2)$ (see Ref. \cite{khodas07}), and 
the frequency dispersions.

Comparison with numerical data by Caux and Calabrese \cite{caux06} (figure \ref{fig:dsf}) 
shows that the suggested formula nicely works in the regimes of both weak and strong 
coupling.

Concluding, we propose the approximate formula (\ref{dsfapp1}) for DSF of the 
one-dimensional Bose gas at zero temperature. It neglects, in effect, only the 
contributions of multiparticle excitations outside the bounds given by the dispersion 
curves $\omega_\pm$, whose contribution is small. Our formula is consistent with 
predictions of Luttinger liquid theory, has the exact exponents at the edge of the 
spectrum, gives the correct first-order expansion in the strong-coupling regime, and 
shows nice agreement with available numerical data.

The authors are grateful to Jean-Sebastien Caux for making the data of numerical 
calculations of Ref.~\cite{caux06} available to us and to Thomas Ernst for checking our 
numerical results. This project received funding from the Marsden fund of New Zealand 
under contract number MAU0607. AYuCh thanks Massey University for hospitality.

\section*{References}

\begin{thebibliography}{10}
\bibitem{note}
We slightly change the notations: ours $\omega_{\pm}$ and $\pm\mu_{\pm}$
  correspond to $\omega_{1,2}$ and $\mu_{1,2}$, respectively, in
  Ref.~\cite{imambekov08}.
\end{thebibliography}
\providecommand{\newblock}{}

\end{document}